\documentclass[aps,prd,twocolumn,superscriptaddress]{revtex4}
\usepackage{graphicx}
\usepackage{color}

\begin{document}
\baselineskip=12pt
\def\black{\textcolor{black}}
\def\red{\textcolor{black}}
\def\blue{\textcolor{blue}}
\def\green{\textcolor{black}}
\def\beq{\begin{equation}}
\def\eeq{\end{equation}}
\def\bea{\begin{eqnarray}}
\def\eea{\end{eqnarray}}
\def\orc{\Omega_{r_c}}
\def\om{\Omega_{\text{m}}}
\def\E{{\rm e}}
\def\bearst{\begin{eqnarray*}}
\def\eearst{\end{eqnarray*}}
\def\peleven{\parbox{11cm}}
\def\peffec{\peight{\bearst\eearst}\hfill\peleven}
\def\pspace{\peight{\bearst\eearst}\hfill}
\def\ptwelve{\parbox{12cm}}
\def\peight{\parbox{8mm}}


\title{Prospects for Detecting Dark Matter Halo Substructure with Pulsar Timing}

MCTP-11-05

\author{Shant Baghram}
\email{baghram@physics.sharif.edu}
\address{Department of Physics, Sharif University of
Technology, P.O.Box 11365--9161, Tehran, Iran}
\author{Niayesh Afshordi}
\address{Perimeter Institute for Theoretical Physics, 31 Caroline St. N.,
Waterloo, ON, N2L 2Y5, Canada}
\address{ Department of Physics and Astronomy, University of Waterloo, 200 University Avenue West,
Waterloo, ON, N2L 3G1, Canada}
\author{Kathryn M. Zurek}
\address{Michigan Center for Theoretical Physics, Department of Physics, University of  Michigan, Ann Arbor, Michigan 48109, USA}
\vskip 1cm

\begin{abstract}
One of the open questions of modern cosmology is the nature and
properties of the dark matter halo and its substructures. In this
work we study the gravitational effect of dark matter substructures
on pulsar timing observations. Since millisecond pulsars are stable
and accurate emitters, they have been proposed  as plausible
astrophysical tools to probe the gravitational effects of dark
matter structures. We study this effect  on pulsar timing through
Shapiro time delay (or integrated Sachs-Wolfe (ISW) effect) and
Doppler effects statistically, showing that the latter dominates the
signal. For this task, we relate the power spectrum of pulsar
frequency change to the matter power spectrum on small scales, which
we compute using the stable clustering hypothesis, as well as other
models of nonlinear structure formation. We compare this power
spectrum with the reach of current and future observations of pulsar
timing designed for gravitational wave detection. Our results show
that while current observations are unable to detect these signals,
the sensitivity of the upcoming square kilometer array is only a
factor of few weaker than our optimistic predictions.
\\

\end{abstract}

\maketitle

\newpage

\section{INTRODUCTION}

One of the greatest puzzles of modern cosmology is the nature of the
dark matter (DM). The latest cosmological observations indicate that
DM has a mean cosmic mass density $\sim$5 times larger than the
density of the baryonic matter ({\em e.g}. see WMAP- 7 yr results
\cite{WMAP}), and its presence is confirmed by a large amount of
astrophysical evidence, such as rotation curves of galaxies,
gravitational lensing effects, growth of large scale structure of
the Universe, big bang nucleosynthesis  and the dynamics of the
Universe as a whole \cite{Roos2010}. The cosmological and
astrophysical observations that may lead us to better understand
this unknown component of the Universe are  areas of intense study,
focusing on DM's nature as particles, its structure, distribution
and effect on the other components of the Universe. Studying the
small scale structure of DM, for example, will tell us something
about the DM particle's fundamental properties.  Consequently,
finding new footprints of DM in astrophysical observations is
important for opening new horizons in DM studies.

The theory of structure formation, which is based on gravitational
instability of primordial matter density fluctuations and the
hierarchal scheme of structure formation, assumes that collisionless
DM is the main ingredient in today's cosmological structures. One of
the features of this theory is that DM collapses into bound states,
known as DM halos. A cold dark matter primordial power spectrum
predicts a large range of mass scales for these DM halos, from
$10^{12}-10^{14}M_{\odot}$ down to $10^{-12}-10^{-4}M_{\odot}$
\cite{kris2006}. Larger halos form from merger of smaller halos
which may partly survive as substructure of bigger halos.

The statistics of DM distribution and the dynamics of this
substructure may have an effect on astrophysical observations. One
of the promising astrophysical probes for studying the distribution
of interstellar medium (ISM) which is considered to be mostly
baryonic matter are pulsars \cite{Frail1994}. ISM causes  a
dispersion on pulsars' light which in turn has an effect on pulsar
timing residuals.

In the present work, we push this one step further and study the
{\it{gravitational effects}} of dark matter halo substructure on
pulsar timing. This effect manifests itself through the (1) Shapiro
time-delay effect \cite{Shapiro1964}, and (2) Doppler effect. The
Shapiro time delay is caused by the presence  of dark matter's
dynamical potential along the line of sight. On the other hand, the
Doppler effect is caused by the acceleration of the observer/pulsar
because of the pull of DM subhalos. Probing dark matter substructure
by Shapiro time delay in pulsar timing was first proposed by Siegel
{\it et al.} by considering the effect of one DM subhalo crossing
the line of sight \cite{Siegel2007}. Dark matter studies  with
pulsar timing  continued by Seto and Cooray \cite{Seto2007},
Pshirkov {\it{et al.}}\cite{Pshirkov2008} and recently by Ishiyama
{\it {et al.}}\cite{Ishiyama2010} . On the other hand the Shapiro
delay effect was studied for relativistic neutrino and photons of
SN1987A \cite{Longo1988} and also for low frequency pulsars  in
globular cluster \cite{Larchenkova2006}. It is worth mentioning that
pulsar timing was also used to study other astronomical effects
\cite{Ohnishi1995}, as a recent example the effect of DM subhalos
crossing the line of sight  was studied in astrometric microlensing
\cite{Erick2010}.

In this work, as a complementary and more realistic view, we
consider the statistical distribution of DM substructure and its
effect  on pulsars' timing residual. In order to study the effect of
the DM substructure distribution on pulsar timing, we need a
structure formation model. On very  small scales deep into the
nonlinear regime of structure formation, which is unaffected by halo
merging or tidal disruption, we can use the stable clustering
hypothesis. The stable clustering hypothesis was first introduced by
Davis and Peebles \cite{Peebles77} as an analytic technique to study
the galaxy correlation function in the deeply nonlinear regime, and
was subsequently applied to fitting formulas for nonlinear
correlation functions/power spectra \cite{Hamilton1977,Peacock1996}.
In the current work, we use the  phase-space stable clustering model
which was recently developed by Afshordi {\it {et al.}}
\cite{Afshordi2009}.

The article is structured as follows.  In Sec.~(II), we first
introduce millisecond pulsars. Then in the following subsections we
derive the power spectrum of frequency change of pulsars for Shapiro
time delay and Doppler effects. In Sec.~(III), we review the stable
clustering hypothesis in phase-space. In Sec.~(IV), we find  the
frequency change power spectrum and show its dependence on free
parameters of the model, both for Shapiro and Doppler effects. In
Sec.~(V), we discuss the observational prospects of detecting these
effects with current and future pulsar timing arrays. Finally,
Sec.~(VI) concludes the paper.

For reference, we set cosmological parameters to be
$\Omega_{m}^{0}=0.27$, $\sigma_{8}=0.8$ and $H_{0}=100h$ km/s/Mpc
where $h=0.7$.

\section{Gravitational effect on Power Spectrum of pulsar timing}
\label{Section2}

In this section, we first introduce millisecond pulsars as
promising astrophysical observational probes to detect the
gravitational effects of DM substructures. Then we derive the
statistics of frequency change due to Shapiro and Doppler effects.

\subsection{Millisecond Pulsars }

The most stable, consistent astrophysical emitters in the known
universe are millisecond pulsars, many of them remaining stable
without flux change over timescales exceeding 30 years
\cite{Verb2009}.
On account of this they have been used as precise tools to probe
changes in the matter distribution between the pulsar and earth
\cite{Jenet2006}. The pulsars with the highest rotational
frequencies, and hence the shortest pulse to pulse periods, are the
most stable with a time period of ${\cal O}(1\mbox{ ms})$. The
typical residual of these pulsars is of order of ${\cal O}(1~\mu
\mbox{s})$. This means that fluctuations in pulsar period within a
short time scale (e.g. $\sim$ 1 hr) are less than $\sim \mu$s. These
residuals  do not accumulate, which means that the period remains
constant during the time that a pulsar is stable. This is used to
measure the  pulsar's timing residuals with high accuracy during a
long period ($\sim 10$ years), and to search for nonintrinsic
changes in pulsar timing. Consequently, to detect  any physics
besides the pulsars' intrinsic changes, we should search for a time
delay larger than the intrinsic uncertainties. An important point is
that many interesting nonintrinsic effects on pulsar timing will be
correlated. An example is the attempt to detect gravitational waves
through cross-correlation of  pulsar timing arrays \cite{Jenet2005}.
Another possible nonintrinsic effect which we consider in this work
is the  change of the gravitational potential. The transit of DM
halo substructure across the line of sight, which causes the Shapiro
delay, is studied in the following subsection.  This discussion is
followed by a consideration of the Doppler effect, caused by the
acceleration of pulsar/observer due to presence of DM substructure.

\subsection{Shapiro time delay}

The Shapiro time delay is caused by the presence of a time dependent
gravitational potential along  the line of sight. To quantify this effect, we can write the  metric of
perturbed space time as
\begin{equation}
ds^2=-(1+2\Phi)dt^2+(1-2\Phi)d\vec{x}^2,
\end{equation}
where $\Phi$  represents the Newtonian potential in the weak field
limit and we set the speed of light $c=1$. In the case of
pulsars, we can write the null geodesics for a pulse received at
time $t$ using the above metric as
\begin{equation}
t=t_{0}+\delta t=\int_{x_{em}}^{x_{obs}}(1-2\Phi)d\vec{x},
\end{equation}
where $\delta t$ is obtained from integration over the perturbed
potential along the  line of sight. As it is not possible to measure
the absolute light travel time  of any astrophysical object, we need
to observe the time arrival changes over a  detection period. The
time derivative of the pulsar time residual is defined as
\begin{equation}
\dot{\delta t}=-\frac{\delta\nu}{\nu}=-2\int \dot{\Phi}d\vec{x},
\end{equation}
where $\nu$ is the frequency of the pulsar, and $\delta\nu$ is the
change in frequency (we note that this is identical to the cosmological
Integrated Sachs-Wolfe (ISW) effect \cite{Sachs1967}). In order  to
find a statistical description of the induced time delays, we assume that DM substructures move with a constant velocity $v$ across the line of sight in the $x$-direction. We refer to this as the {\it moving screen} approximation, which enables us to relate the time derivatives to the spatial gradients of metric, {\em i.e.}  $\dot{\Phi}=v\frac{\partial}{\partial
x}\Phi$. The temporal correlation of the frequency changes is then given by
\begin{equation}\label{corr1}
\langle(\frac{\delta\nu}{\nu})_{{\i}}(\frac{\delta\nu}{\nu})_{{\i}{\i}}\rangle=4\int_{0}^{z_{0}}\int_{0}^{z_{0}}dz_{{\i}}dz_{{\i}{\i}}v^2\langle\frac{\partial}{\partial
x}\Phi_{{\i}}\frac{\partial}{\partial_{x}}\Phi_{{\i}{\i}}\rangle,
\end{equation}
where $z_{0}$ is the position of the pulsar in the $z$-direction (line of
sight), which we take to be $\sim$ 1 kpc, and
$\Phi_{{\i}}$ and $\Phi_{{\i}{\i}}$ correspond to potentials at two
different times.  Now we can express the right hand side of Eq.~(\ref{corr1})
in Fourier space as:
\begin{eqnarray}\label{corr2}
\langle(\frac{\delta\nu}{\nu})_{{\i}}(\frac{\delta\nu}{\nu})_{{\i}{\i}}\rangle=4v^2\int_{0}^{z_{0}}\int_{0}^{z_{0}}dz_{{\i}}dz_{{\i}{\i}}\int
\frac{d^3\vec{k_{{\i}}}}{(2\pi)^3}\int\frac{d^3\vec{k_{{\i}{\i}}}}{(2\pi)^3}\\
\nonumber
(ik^{x}_{{\i}})(ik^{x}_{{\i}{\i}})\langle\Phi(\vec{k_{{\i}}})\Phi(\vec{k_{{\i}{\i}}})\rangle
e^{-i\vec{k}_{{\i}}.\vec{r}_{{\i}}}e
^{-i\vec{k_{{\i}{\i}}}.\vec{r}_{{\i}{\i}}}.
\end{eqnarray}
By integrating over $z_{{\i}}$ and $z_{{\i}{\i}}$, and using the
definition of the potential power-spectrum,
\begin{equation}
\langle\Phi(\vec{k}_{{\i}})\Phi(\vec{k}_{{\i}})\rangle=(2\pi)^3\delta^3(\vec{k}_{{\i}}+\vec{k}_{{\i}})P_{\Phi}(\vec{k}),
\end{equation}
Eq.~(\ref{corr2}) becomes
\begin{eqnarray}
\langle(\frac{\delta\nu}{\nu})_{{\i}}(\frac{\delta\nu}{\nu})_{{\i}{\i}}\rangle&=&4v^2\int\frac{d^3k_{{\i}}}{(2\pi)^3}\left[k^{x}_{{\i}}{z_{0}{\rm sinc}\left(\frac{k^{z}_{{\i}}z_{0}}{2}\right)}\right]^2\\
\nonumber
&\times&P_{\Phi}(\vec{k_{{\i}}})e^{-i{k}_{{\i}}^{x}(x_{{\i}}-x_{{\i}{\i}})},
\end{eqnarray}
where ${\rm sinc}(x)\equiv \frac{\sin(x)}{x} $. We can take the integral over $k^{z}$,
which results in
\begin{eqnarray}{\label{EqintTime}}
\langle(\frac{\delta\nu}{\nu})_{{\i}}(\frac{\delta\nu}{\nu})_{{\i}{\i}}\rangle&=&4z_{0}v^2\int\int\frac{dk_{y}}{2\pi}\frac{dk_{x}}{2\pi}P_{\phi}(\vec
k)\\ \nonumber &\times&(k_{x})^2e^{-ik^{x}v(t_{1}-t_{2})},
\end{eqnarray}
 where we omit the subscript ${\i}$ and replace $x_{\i}=vt_{\i}$
in the moving screen approximation (we also assume $k_{z}\sim
z_{0}^{-1} \ll k_{x}, k_{y}$). Now by considering the definition of
the time-delay power-spectrum,
\begin{equation}\label{EqPS}
\langle(\frac{\delta\nu}{\nu})_{{\i}}(\frac{\delta\nu}{\nu})_{{\i}{\i}}\rangle=\frac{1}{2\pi}\int P_{\frac{\delta\nu}{\nu}}(\omega)e^{-i\omega\Delta t}d\omega,
\end{equation}
we can integrate  Eq.~({\ref{EqintTime}}) over the time difference
of two observations to obtain the power-spectrum,
\begin{eqnarray}\label{EqintTime2s}
&&P_{\frac{\delta\nu}{\nu}}(\omega)=\int\langle(\frac{\delta\nu}{\nu})_{{\i}}(\frac{\delta\nu}{\nu})_{{\i}{\i}}\rangle
e^{i\omega\Delta t}d(\Delta t)\\ \nonumber &&=\int d(\Delta
t)4z_{0}v^2\int\int\frac{dk_{y}}{2\pi}\frac{dk_{x}}{2\pi}P_{\phi}(\vec
k)(k_{x})^2e^{-ik^{x}v(\Delta t)}e^{i\omega\Delta t},
\end{eqnarray}
where $\Delta t=t_{1}-t_{2}$. Now the integration over $dk_{x}$ and
$d(\Delta t)$ gives us the relation between $k_x$ and the frequency as $k_{x}v=\omega$. Consequently
Eq.~(\ref{EqintTime2s}) simplifies to:
\begin{equation}\label{corr3}
\omega P_{\frac{\delta\nu}{\nu}}(\omega)=\frac{4z_{0}}{
v}\int\frac{dk^{y}}{2\pi}\omega^3P_{\phi}\left(\sqrt{\frac{\omega^2}{v^2}+k_{y}^2}.
\right)
\end{equation}
Using the Poisson equation, we can relate the potential
power-spectrum to the matter power-spectrum $P_{\rho}(\vec{k})$:
\begin{equation}\label{EqPhi}
P_{\Phi}(\vec{k})=\left(\frac{4\pi
G}{k^2}\right)^2P_{\rho}(\vec{k}).
\end{equation}
 Finally by
inserting Eq.~(\ref{EqPhi}) in Eq.~(\ref{corr3}), we find the
dimensionless $\omega P(\omega)$ in terms of the matter
power-spectrum: \bea\label{powershapiro} &&\omega
P_{\frac{\delta\nu}{\nu}}(\omega)|_{_{\rm
Shapiro}}=\nonumber\\&&\frac{4z_{0}}{
v}\int\frac{dk_{y}}{2\pi}\omega^3\left(\frac{4 \pi
G}{k^2}\right)^2{\bar{\rho}}^2P_{NL}\left(\sqrt{\frac{\omega^2}{v^2}+k_{y}^2}\right),
\eea where we replace
$P_{\rho}(\vec{k})={\bar{\rho}}^2P_{NL}(\vec{k})$ ,
in which $\bar{\rho}$ is the mean cosmic DM density. 
In Sec.(IV), we will
derive this function by using the stable clustering hypothesis.

\subsection{Doppler effect}

Changing the potential of  DM substructure near  pulsars or the
Earth will introduce a velocity shift, which affects pulsar
frequencies via the Doppler effect. In the Doppler effect, the
frequency change of a pulsar is related to the line of sight
velocity as $\frac{\delta\nu}{\nu}=v_{l.s.}$. So the correlation of
the frequency changes observed at two separate times
 $t_{1}$ and $t_{2}$ caused by the Doppler effect can be written as
\begin{eqnarray}\label{dopcor1}
&&\langle(\frac{\delta\nu}{\nu})_{{\i}}(\frac{\delta\nu}{\nu})_{{\i}{\i}}\rangle=\int_{-\infty}^{t_{1}}dt\int_{-\infty}^{t_{2}}dt^\prime\langle\nabla_{z}\Phi_{{\i}}\nabla_{z}\Phi_{{\i}{\i}}\rangle\\
\nonumber &\times&
e^{\varepsilon(t-t_{1})}e^{\varepsilon(t^\prime-t_{2})},
\end{eqnarray}
where $\varepsilon$ is a small parameter to regulate the infrared
divergence of the integral.
Once more, we can write the right hand side of Eq.~(\ref{dopcor1}) in
Fourier space. Integration  over time variables with the limit of
$\varepsilon\rightarrow 0$ results in
\begin{equation}{\label{dopcor2}}
\langle(\frac{\delta\nu}{\nu})_{{\i}}(\frac{\delta\nu}{\nu})_{{\i}{\i}}\rangle=
\int\frac{d^3\vec{k}}{(2\pi)^3}P_{\phi}(\vec
k)\frac{(k_{z})^2}{(k_{x}v)^2}e^{-ik^{x}v(\Delta t)},
\end{equation}
where we used the moving screen approximation to replace time
integrals by integrals over $x$. Again, using Eq.~(\ref{EqPS}), we
can write the power-spectrum of frequency changes as
\begin{eqnarray}{\label{EqintTime2}}
&&P_{\frac{\delta\nu}{\nu}}(\omega)=\int\langle(\frac{\delta\nu}{\nu})_{{\i}}(\frac{\delta\nu}{\nu})_{{\i}{\i}}\rangle
e^{i\omega\Delta t}d(\Delta t)\\ \nonumber&=&\int d(\Delta t)
\int\frac{d^3\vec{k}}{(2\pi)^3}P_{\phi}(\vec
k)\frac{(k_{z})^2}{(k_{x}v)^2}e^{-ik^{x}v(\Delta t)}e^{i\omega\Delta
t}.
\end{eqnarray}
Integration over $dk_{x}$ and $d\Delta t$ gives us the
relation between $k_x$ and the frequency as
$\omega=k_{x}v$. Consequently Eq.~(\ref{EqintTime2}) results in
\begin{equation}\label{EqintTime3}
P_{\frac{\delta\nu}{\nu}}(\omega)=\frac{1}{v} \int\frac{dk_{y}}{2\pi}\int\frac{dk_{z}}{2\pi}\frac{k_{z}^2}{\omega^2}P_{\Phi}(\vec{k}).
\end{equation}
Because of the symmetry between the integration over $k_{y}$ and
$k_{z}$, we can replace $k^2_y$ by $(k^2_y+k^2_z)/2$ in Eq. (\ref{EqintTime3}), which leads to
\begin{equation}
P_{\frac{\delta\nu}{\nu}}(\omega)=\frac{1}{4\pi} \int dk_{*}
\frac{k_{*}^3}{v\omega^2}P_{\phi}
\left(\sqrt{\frac{\omega^2}{v^2}+k_{*}^2} \right),
\end{equation}
where $k_{*}=\sqrt{k_{y}^2+k_{z}^2}$.  By using the Poisson
equation, we can relate the potential power spectrum to matter power
spectrum, and finally write the dimensionless power spectrum as
\bea\label{powerdoppler}
&&\omega P_{\frac{\delta\nu}{\nu}}(\omega)|_{_{\rm Doppler}}=\nonumber\\&&\frac{1}{4\pi} \int
dk_{*} \frac{k_{*}^3}{v\omega}\left(\frac{4\pi
G}{k^2}\right)^2{\bar{\rho}}^2P_{NL}\left(\sqrt{\frac{\omega^2}{v^2}+k_{*}^2}
\right).
\eea
As in the case of Shapiro delay,  by knowing the matter power
spectrum we can determine the pulsar timing power spectrum.

\section{STABLE CLUSTERING HYPOTHESIS}
\label{Section3} In this section, we discuss the stable clustering
hypothesis as a model to describe the nonlinear structure formation
that will give us the required matter power spectrum. We make use of
the phase-space stable clustering model recently developed  by
Afshordi {\it {et al.}} \cite{Afshordi2009}. The collisionless
Boltzmann equation at the phase-space coordinates, ${{\vec{r}+\Delta
\vec{r}, \vec{v}+\Delta \vec{v}}}$ is approximately given by
\begin{eqnarray}\label{Boltz}
&&\frac{df}{dt}(r+\Delta r,v+\Delta v, t) \simeq \\ \nonumber
&&\frac{\partial f}{\partial t}+\frac{\partial f}{\partial
r}\cdot(v+\Delta v)-\frac{\partial f}{\partial
v}\cdot\left[\nabla\Phi+(\Delta r\cdot\nabla)\nabla\Phi\right]=0,
\end{eqnarray}
where $\Phi$ is the gravitational potential, and for simplicity we
omit the vector signs of distances and velocities. We can reexpress
the above equation in terms of the  phase-space density in the
comoving coordinates with particle $i$:
\begin{equation}
\tilde{f}_{i}(\Delta r,\Delta v)\equiv f(r_{i}+\Delta r, v_{i}+\Delta v).
\end{equation}
Using this new function, we can write the Boltzman Eq.~(\ref{Boltz})
as
\begin{equation}\label{Boltz1}
\frac{df}{dt}=\frac{\partial\tilde{f}_{i}}{\partial t}|_{\Delta
r,\Delta v}+\frac{\partial\tilde{f}_{i}}{\partial\Delta r}\cdot\Delta
v-\frac{\partial\tilde{f}_{i}}{\partial\Delta
v}\cdot(\Delta r\cdot\nabla)\nabla\Phi=0.
\end{equation}
Notice that Eq. (\ref{Boltz1}) can be understood as the tidal limit of the Boltzmann equation in terms of the phase coordinates $(\Delta r,\Delta v)$, {\em i.e.} in the coordinate system comoving with particle $i$.


\begin{figure}
\includegraphics[width=1.2\linewidth]{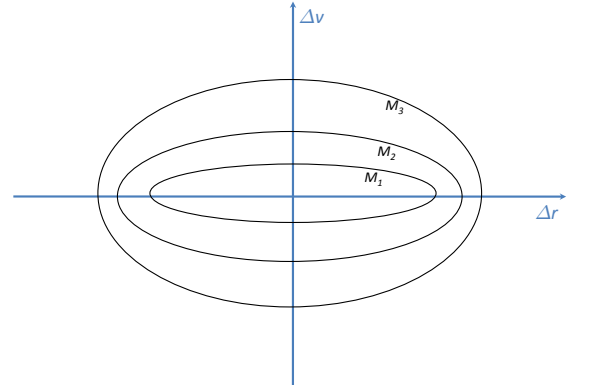}
\caption {Surfaces of constant average CDM phase space density, $\langle\tilde{f}\rangle_{p} = \mu \xi_s$, around a typical particle in the stable clustering hypothesis. The surfaces are assumed to be concentric ellipsoids (Eq. \ref{eq:f_tilde_p}). The mass and Hubble scales at the collapse of the structure,  $M(\xi_s)$ and $H(\xi_s)$, are related to the phase space density, $\mu \xi_s$, on each surface via the spherical collapse results Eqs. (\ref{spcoll}-\ref{sigmaM}), while $\mu \sim 3$\% is an empirical factor that quantifies tidal stripping and is fixed through comparison with numerical simulations \cite{Afshordi2009}.} \label{stable_pic}
\end{figure}

The stable clustering hypothesis assumes that
$\frac{\partial\tilde{f}_{i}}{\partial t}|_{\Delta r,\Delta v}$
averaged over the particles vanishes for small $\Delta r$ and
$\Delta v$. This implies that the number of neighbors within a fixed physical separation of a DM particle in the phase space does not vary with time. Now, if we assume that ${\langle\tilde{f}_{i}\nabla\nabla
\Phi\rangle}_{p}\approx\langle\tilde{f}_{i}\rangle_{p}\langle\nabla\nabla\Phi\rangle_{p}$,
then a solution to Eq.~(\ref{Boltz1}) is:
\begin{equation}\label{fp}
\langle\tilde{f}\rangle_{p}\equiv\frac{1}{N}\sum_{i}\tilde{f}_{i}=F[\Delta
v^2+\Delta x_{j}\Delta
x_{k}\langle\partial_{j}\partial_{k}\Phi\rangle_{p}],
\end{equation}
where $F$ is the general solution with isotropic velocity
distribution and $N$ is the number of particles in the phase space
volume of interest. By using the approximation of a spherically
symmetric potential, the above solution can be rewritten  by
applying the Poisson equation:
\begin{equation}
\langle\tilde{f}\rangle_{p}=\mu\xi_{s}=F[(\Delta v)^2+100H^2(\xi_{s})(\Delta r)^2],\label{eq:f_tilde_p}
\end{equation}
where $\xi_{s}$ and $H(\xi_s)$ are the phase-space density and Hubble constant at the formation time of
DM substructure, respectively (see Fig. \ref{stable_pic}). We also use  the spherical collapse model prediction for the halo density, which is roughly $~$200 times the critical
density at the formation time \cite{Gunn72}. $\mu \simeq 3\%$ is
the mean fraction of bound {\it particle pairs} that can survive the tidal
disruption period, and is calibrated by comparison with N-body simulations \cite{Afshordi2009}. To determine the function $F$, we use the
spherical collapse model results. The phase-space density can be
expressed as
\begin{equation} \label{spcoll}
\xi_{s}\sim\frac{10H(\xi_{s})}{G^2M(\xi_{s})},
\end{equation}
using the fact that the radius and velocity dispersion of halos are
related as
\begin{equation}
\sigma_{vir}\sim10Hr_{vir}.
\end{equation}
The phase-space volume of the collapsed halo, {\em i.e.} the volume
of the constant-$\xi_{s}$ ellipsoid in Eq.~(\ref{fp}), is
$M/\xi_{s}$, and by using Eq.~(\ref{spcoll}), we find
\begin{equation}\label{minusF}
\left[\frac{\pi
F^{-1}(\mu\xi_{s})}{10H(\xi_{s})}\right]^3=\frac{[GM(\xi_{s})]^2}{10H(\xi_{s})}.
\end{equation}
Furthermore, the mass scale that collapses at a given cosmological
epoch is characterized by
\begin{equation}\label{sigmaM}
\left[\frac{H(\xi_{s})}{H_{0}}\right]^{-2/3}\sigma[M(\xi_{s})]\sim\delta_{c}\simeq1.7,
\end{equation}
where $\delta_{c}$ is the linear density threshold for the spherical
collapse, $\sigma[M]$ is the rms top-hat linear over density at the
mass scale $M$, and $H_{0}$ is the Hubble constant in the present
epoch. Using the above result, the phase-space correlation function
is obtained as
\begin{eqnarray}\label{eq:boltz-cor}
&&\langle f(r_{1},v_{1})f(r_{2},v_{2})\rangle \nonumber\\
&\simeq&\frac{1}{V_{6}}\int_{V_{6}}d^{3}rd^{3}v f(r,v)f(r+\triangle
r,v+\triangle v) \nonumber\\ &=&\frac{1}{V_{6}}\sum_{i}
f(r_{i}+\triangle r,v_{i}+\triangle
v)=\frac{N}{V_{6}}\langle\tilde{f}\rangle_{p}\nonumber \\
&\simeq&\langle f(r_{1},v_{1})\rangle\langle
f(r_{2},v_{2})\rangle+\mu\langle
f(\bar{r},\bar{v})\rangle\xi_{s}(\triangle r,\triangle v).\nonumber\\\label{ff}
\end{eqnarray}
In the equation above we used the assumption of ergodicity to
replace the ensemble  average $\langle\rangle$  by the volume
average, in a given volume of phase-space $V_{6}$, while
$(\bar{r},\bar{v})$ are the mean values of $(r_{1},v_{1})$ and
$(r_{2},v_{2})$. The second term is based on the stable clustering
described above, with the assumption that $|\triangle
v|=|v_{1}-v_{2}|\ll\triangle v_{tid}$ and $|\triangle
r|=|r_{1}-r_{2}|\ll\triangle r_{tid}$ where $\triangle v_{tid}$ and
$\triangle r_{tid}$ characterize the tidal truncation radii in the
phase-space. On the other hand, the first term in Eq.~(\ref{ff})
dominates for large separations in the phase-space, where particles
are not correlated. So Eq.~(\ref{ff}) is an interpolation between
the stable clustering and the smooth halo regimes. This is a crucial
point in calculating the nonlinear power spectrum  of structures on
small scales where it is  related to phase-space density correlation
$\mu\langle f(\bar{r},\bar{v})\rangle\xi_{s}(\triangle r,\triangle
v)$ term.


\section{Pulsar Residual Power Spectrum from stable clustering hypothesis}
\label{Section4}

\begin{figure}
\includegraphics[width=\linewidth]{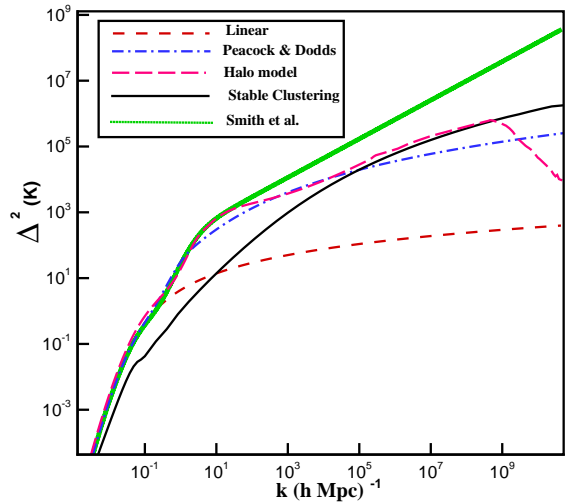}
\caption {Dimensionless power spectrum of density fluctuations
$\Delta^2(k)=\frac{k^3P_{NL}(k)}{2\pi^2}$ as a function of wavenumber
$k$ for the linear regime (short-dashed line), Peacock and Dodds fitting
formula (dash-dotted line), Smith et al. fitting formula (thick dotted line), halo model (long-dashed line)
and for $k \lesssim 10^2$stable clustering hypothesis used in this work (solid line).} \label{power-9}
\end{figure}

In order to calculate the dimensionless power spectrum of pulsar
frequency change, we need to know the power spectrum of matter on
small scales.  We now make use of the stable clustering
hypothesis prediction, as developed in the previous section.

To use the stable clustering formula obtained in Eq. (\ref{ff}), we
must relate the matter density  power spectrum in
Eqs.~({\ref{powershapiro},\ref{powerdoppler}}) to the real space
correlation function of densities. In the stable clustering
hypothesis in phase-space, on small scales this relation becomes
\begin{eqnarray} {\label{Eqcorr}}
\langle\rho(\vec{r}_{{\i}})\rho(\vec{r}_{{\i}{\i}})\rangle&=&\int
d^3\vec{v_{{\i}}}d^3\vec{v_{{\i}{\i}}}\langle f(\vec{r}_{{\i}},\vec{v}_{{\i}})f(\vec{r}_{{\i}{\i}},\vec{v}_{{\i}{\i}})\rangle\\
\nonumber &\simeq&\int d^3\bar{\vec{v}}d^3\Delta \vec{v} \mu \langle
f(\vec{\bar{r}},\vec{\bar{v}})\rangle\xi_{s}(\Delta r,\Delta
v)\\ \nonumber &=&\mu\bar{\rho}_{avg}\int
d^3\Delta\vec{{v}}\xi_{s}(\Delta {r},\Delta v).
\end{eqnarray}
In order to find the dependence of the dimensionless power spectrum
$\omega P_{\frac{\delta\nu}{\nu}}(\omega)$ on $\omega$, we should
have the rms top-hat linear overdensity $\sigma(M)$. $\sigma(M)$ is
the integral of linear matter power spectrum on a chosen window
function as
\begin{equation}\label{sigma}
\sigma^2(M)=\int \frac{d^3k}{(2\pi)^3}P_{L}(k)W^2(kR),
\end{equation}
where $P_{L}(k)$ and $W(kR)$ are the linear matter power spectrum
and the Fourier transform of the spherical top-hat filter of radius
$R$, respectively, where
\begin{equation}
P_{L}(k)=Ak^{n_{s}}T^2(k),
\end{equation}
\begin{equation}
W(x)=\frac{3(\sin x-x\cos x)}{x^3}.
\end{equation}
Here $n_{s}$ is the scalar spectral index of primordial matter power
spectrum and $M=4\pi R^3\bar{\rho}_{m}/3$. The transfer
function can be approximated by the BBKS \cite{Bardeen1986} fitting
formula,
\begin{eqnarray}
&&T(k=q\Omega_{m}h^2Mpc^{-1})\approx \frac{\ln[1+2.34q]}{2.34q}\\
\nonumber &\times&[1+3.89q+(16.2q)^2+(5.47q)^3+(6.71q)^4]^{-1/4}.
\end{eqnarray}

Using Eqs.~(\ref{Eqcorr},\ref{sigma}), and the stable clustering
hypothesis, we can find an expression for the matter power spectrum
on small scales, shown in Fig.(\ref{power-9}) for a mass range
$10^{-6}$ to $10^{12}$ solar masses and $\mu=0.03$. For qualitative
comparison, \red{ we also plot the dimensionless power spectrum
$\Delta^2(k)=\frac{k^3}{2\pi^2}P_{NL}(k)$ obtained from the halo
model of structure formation \cite{Cooray:2002dia}, as well as the
fitting formulas of Peacock and Dodds {\cite{Peacock1996}} and Smith
{\it  et al.} for the nonlinear
 power spectrum {\cite{Smith2003}}.} We note that these approximations are based on fits to numerical simulations at $k \lesssim 10^2$ Mpc$^{-1}$,
 while the stable clustering hypothesis (which goes into second term in Eq.(\ref{eq:boltz-cor})) is expected to hold for $k \gg 10$ Mpc$^{-1}$,
 and thus should be a more appropriate measure of small scale dark matter structures.
  On larger scales, the matter power spectrum is dominated by the first term on
 Eq.(\ref{eq:boltz-cor}), which is equivalent to the standard halo model (i.e. the long-dashed line in Fig.(\ref{power-9})).
 The cut-off in the halo model power spectrum is related to the size of smallest halo mass of $M_{min} = 10^{-6} M_{\odot}$.

    Now, using the nonlinear power spectrum obtained from stable
clustering, the dimensionless power spectrum for the Shapiro time-
delay effect can be written as
\begin{widetext}
\beq
\omega P_{\frac{\delta\nu}{\nu}}(\omega)|_{_{\rm Shapiro}}=\frac{4z_{0}}{
v}\mu\bar{\rho}_{halo}\int\frac{dk_{y}}{2\pi}\omega^{3}\frac{(4\pi
G)^2}{k^4}  \times \int4\pi(\Delta r)^2d(\Delta
r)\frac{\sin(k\Delta r)}{k\Delta r}\int d(\Delta v)4\pi (\Delta v)^2
\frac{10H[\xi_s(\Delta r,\Delta v)]}{G^2M[\xi_s(\Delta r,\Delta v)]}.\label{Shapiro1}
\eeq

In the case of the Doppler effect, we can also  find the
dimensionless power spectrum of pulsar frequency change in terms of
phase-space density derived from the stable clustering hypothesis.
In this case Eq.~(\ref{powerdoppler}) is expressed  as
%
\beq
\omega
P_{\frac{\delta\nu}{\nu}}(\omega)|_{_{\rm Doppler}}=\frac{\mu\bar{\rho}_{halo}}{
v}\int\frac{dk_{y}}{4\pi}\frac{k_{y}^3}{\omega}\frac{(4\pi
G)^2}{k^4}  \times\int4\pi(\Delta r)^2d(\Delta
r)\frac{\sin(k\Delta r)}{k\Delta r}\int d(\Delta v)4\pi (\Delta v)^2
\frac{10H[\xi_s(\Delta r,\Delta v)]}{G^2M[\xi_s(\Delta r,\Delta v)]}.
\label{Doppler1}\eeq
\end{widetext}
Notice that $k=\sqrt{(\frac{\omega}{v})^2+k^2_{y}}$, and the Hubble
parameter and the mass are related by $\sigma(M)$ through Eq.~(\ref{sigmaM}).
$\bar{\rho}_{halo}$ is the smoothed halo local density at solar system
which is assumed to be $\sim 10^5 \rho_{crit}$.

\begin{figure}
\includegraphics[width=\linewidth]{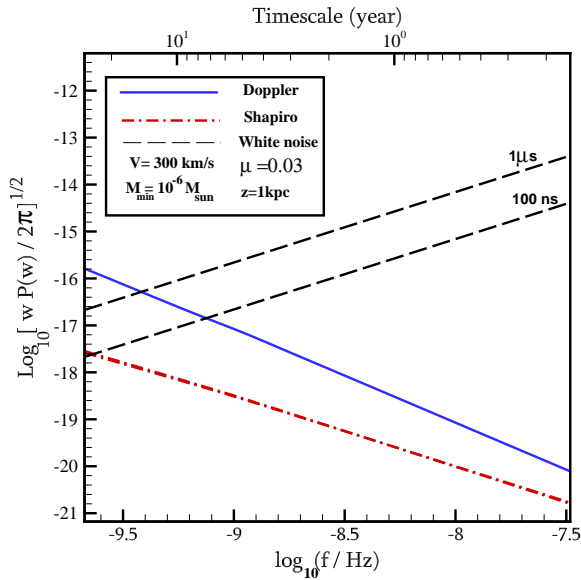}
\caption {Pulsar residual power spectrum as a function of frequency
(bottom x-axis) and the span of observation time in years (top
x-axis) for time delay caused by the Doppler effect (solid line) and
time delay caused by Shapiro effect (dash-dot line). The long dashed
lines represent levels of white noise for 100 ns (bottom) and 1
$\mu$s (top) measured biweekly (see the Appendix) \cite{Verb2009}.}
\label{H-Sh-D}
\end{figure}

In order to numerically perform the third integrations in
Eqs.~(\ref{Shapiro1}-\ref{Doppler1}) over $(k_y,\Delta r,\Delta v)$,
we trade $\Delta v$ with $M$ as the integration variable using
$F^{-1}=(\Delta v)^2+100H^2(\Delta r)^2$ (see
Fig.~\ref{stable_pic}). We then perform the integration in three
steps:
\begin{enumerate}
\item Noting that fixing $M$ and $\Delta r$ fixes $\Delta v$ through Eqs.~(\ref{eq:f_tilde_p}-\ref{sigmaM}), we can first perform the $k_y$ integral for fixed $\Delta r$ and $\Delta v$. Since the integrand can have fast oscillations in $k_y$, we find asymptotic expansions for the $k_y$ integral in the $\omega \Delta r/v \gg$ and $\ll 1$ limit, and devise an interpolation between the two regimes with less than 1\% error, compared to the exact integral.
\item We then perform the $\Delta r$ integral from zero to the maximum of $(F^{-1})^{1/2}/(10H)$, which is fixed by mass $M$, through Eq. (\ref{eq:f_tilde_p}).
\item Finally, we take the integral over substructure mass, $M$, from $M_{min}$ to $M_{max}$, which we  discuss below. {We note that $H[\xi_s(\Delta r,\Delta v)]$ also becomes a function of $M$}.
\end{enumerate}
 Now we are able to calculate
the dimensionless power spectrum $\omega
P_{\frac{\delta\nu}{\nu}}(\omega)$ in terms of the frequency $\omega$
numerically, for Shapiro time delay and the Doppler effect.
For convenience we define the dimensionless parameter $h_{p}$ as
\begin{equation}
h_{p}\equiv[\frac{1}{2\pi}\omega P(\omega)]^{\frac{1}{2}},
\end{equation}
which is shown in Fig.(\ref{H-Sh-D}) for the Doppler effect (solid
line), Shapiro delay (dot-dashed line), and white noise
(dashed line, which is computed in Appendix {\ref{app1}}). In order to calculate  $h_{p}$ numerically, we
consider a realistic set of parameters (but later study the effect
of changing these parameters). We choose the velocity of dark matter
substructures $v=300$ km/s (typical of relative velocities in the Milky Way halo), the typical distance of pulsars to
$z_{0}=1$ kpc, the mean fraction of bound particles that can survive
the tidal disruption period $\mu=0.03$ \cite{Afshordi2009},  the
minimum mass of DM substructure $M_{min}=10^{-6}M_{\odot}$ and also
the maximum $M_{max}=10^{12}M_{\odot}$ (the total mass of a galactic halo).
\red{Later we will show that $h_{p}$ is almost independent of $M_{max}$.}

The power spectra of Shapiro and Doppler effects in
Fig.(\ref{H-Sh-D}) are well described by power-laws:
\beq
\omega P_{\frac{\delta\nu}{\nu}}(\omega)|_{_{\rm Shapiro}} \propto \omega^{-3},~~
\omega P_{\frac{\delta\nu}{\nu}}(\omega)|_{_{\rm Doppler}} \propto \omega^{-4}.\label{eq:scaling}
\eeq
These behaviors can be understood by noticing that the $\Delta v$ integral ({\em i.e.} the last integral) in Eqs.~(\ref{Shapiro1}-\ref{Doppler1}) scales as $H^2(\xi_s)$, if we use the spherical collapse relations of Sec. (\ref{Section3}). Since most small structures with CDM initial conditions collapse around the same time, this is approximately constant. The contribution to the rest of the integrals is dominated by $k^{-1}_y \sim \Delta r \sim v/\omega$, so  the integral over  distances scales as $(\Delta r)^{3} \propto
\omega^{-3}$. Plugging this into Eqs. (\ref{Shapiro1}-\ref{Doppler1}) yields the scalings of Eq. (\ref{eq:scaling}).

{To physically understand the scaling for the Doppler effect we can
once more Fourier transform the power-spectrum in
Eq.~(\ref{Doppler1})  to find that
$v_{Dop.}\sim\frac{\delta\nu}{\nu}$ is proportional to $vt^2 =
(vt)\times t$, {\em i.e.} the magnitude of acceleration is
proportional to distance traveled by the earth/pulsar. This is
exactly what one expects for the gravitational field in a medium
with roughly uniform density, and is due to the fact that most small
substructure forms at roughly the same density $\propto H^2(\xi_s)$.
However, the direction of acceleration is random, as different
substructures will dominate the local gravity on different scales.}

\begin{figure}
\includegraphics[width=\linewidth]{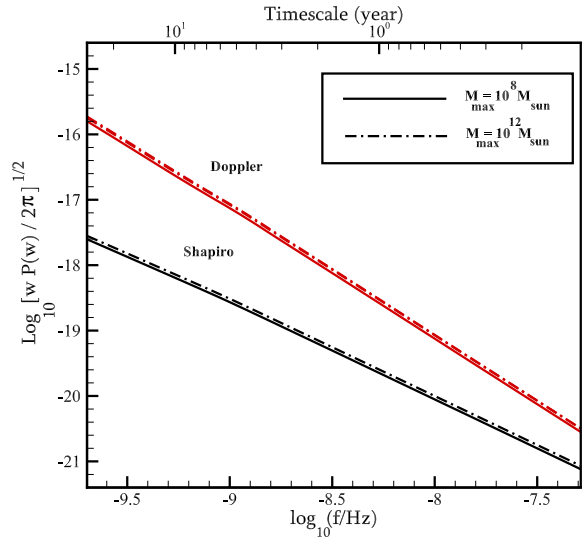}
\caption {Pulsar residual power spectrum as a function of frequency (bottom-x
axis) and the span of time in years (top x-axis)
for time delay caused by the Doppler effect (top-line) and time delay
caused by Shapiro effect (bottom line) for a maximum mass of
halo $M_{max}=10^{12}M_{\odot}$ (dash-dot line) and
$M_{max}=10^8M_{\odot}$ (solid line).}  \label{H-M-max-8}
\end{figure}

An important point to consider before examining the effect of
different parameters on pulsar timing is the study of the effect of
maximum mass in the integrals. As we show in
Fig.~({\ref{H-M-max-8}}), the total dependence of $h_{p}$ on maximum
mass is small, where we plot the $h_{p}$ for
$M_{max}=10^{12}M_{\odot}$, the total mass of a typical galaxy and
$M_{max}=10^8M_{\odot}$, for a more realistic tidal cut-off for
subhaloes at our position in the Milky Way. This confirms that, not
surprisingly, most of the observable effects on pulsar timing comes
from CDM small scale structure.

\begin{figure}
\includegraphics[width=\linewidth]{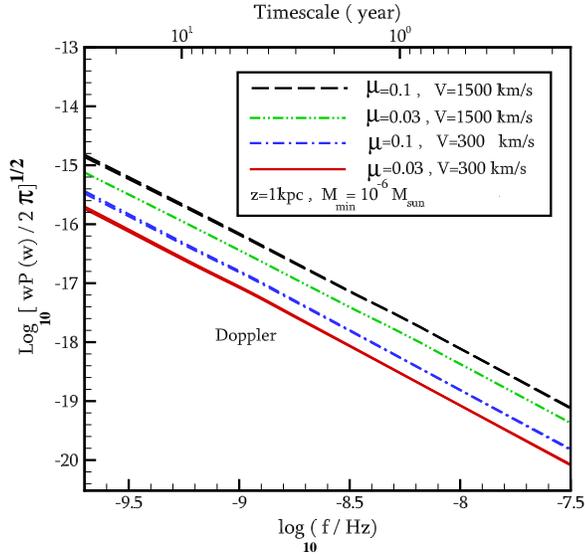}
\caption{Pulsar residual power spectrum as a function of frequency (bottom-x
axis) and the span of time in years (top x-axis)
for Doppler effect (solid line) for different velocities and $\mu$ of
dark matter substructures.} \label{H-D-2}
\end{figure}

 Now we examine the dependence of the power spectrum on different
parameters of the model. We plot the dimensionless amplitude
$h_{p}$ for the Doppler effect for different velocities of dark matter
substructures and the $\mu$-parameter of stable clustering in
Fig.~(\ref{H-D-2}), which shows that  $h_{p}$ is proportional to
velocity and the square root of the $\mu$ parameter.

\begin{figure}
\includegraphics[width=\linewidth]{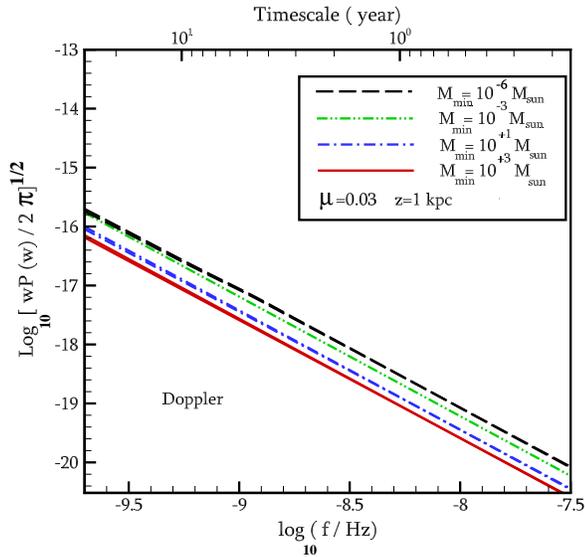}
\caption{Pulsar residual power spectrum as a function of frequency (bottom-x
axis) and the span of time in years (top x-axis)
for Doppler effect (solid line) for different minimum masses
of dark matter substructures.} \label{H-D-M-3}
\end{figure}

In Fig.~(\ref{H-D-M-3}), we plot the power spectrum for different
mass minima of DM substructures. As shown in
Fig.~(\ref{H-D-M-3}), the $\omega^{-4}$ dependence of $h^2_p$
does not change by changing  the minimum of the mass.  However, the
amplitude of the signal increases when the interval of integration
is increased.

\begin{figure}
\includegraphics[width=\linewidth]{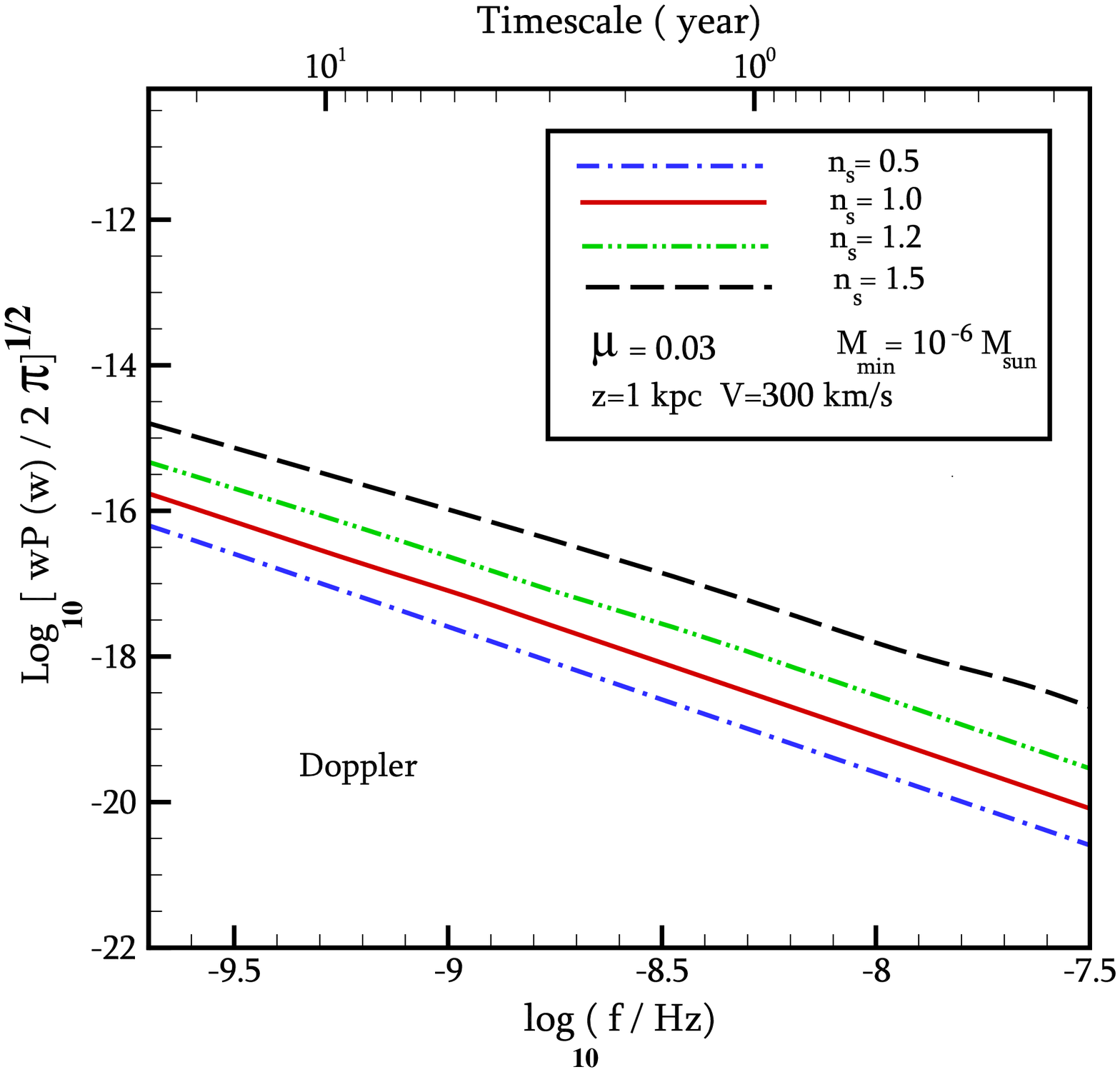}
\caption{Pulsar residual power spectrum as a function of frequency (bottom-x
axis) and the span of time in years (top x-axis)
for Doppler effect for different primodial index of matter power
spectrum.} \label{H-D-n-4}
\end{figure}

\begin{figure}
\includegraphics[width=\linewidth]{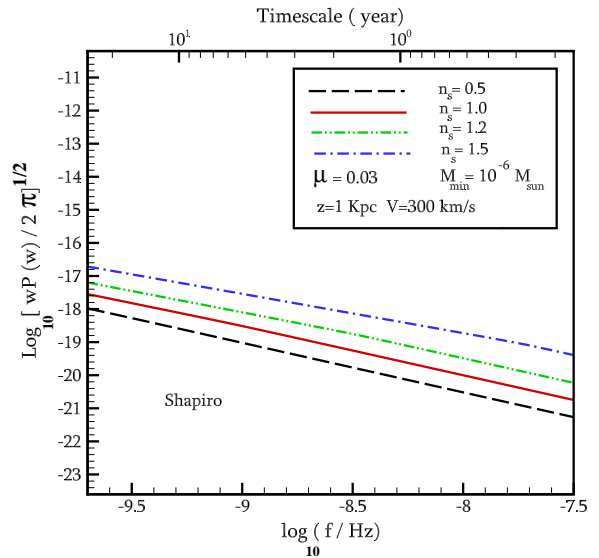}
\caption{Pulsar residual power spectrum as a function of frequency (bottom-x
axis) and the span of time in years (top x-axis)
for Shapiro effect for different primordial index of matter power
spectrum.} \label{H-Sh-n-5}
\end{figure}

In  Figs.~($\ref{H-D-n-4}$) and ($\ref{H-Sh-n-5}$) we plot $h_{p}$ given
different primordial spectral index $n_{s}$, for Doppler and Shapiro
effects respectively. For $n_{s}<1$, the slope of $h_{p}$ does not
change, as $\sigma(M)$ becomes flat for low masses. On the other
hand for larger $n_{s}$, we see a shallower  $\omega$ dependence for
$h_{p}$ as there is more power on small scales.

{\section{Observational Prospects}}

Finally, to explore the observational prospects  for the detection of pulsar frequency change due to
dark matter substructures, we  compare our results with the observational bounds put
on detection of gravitational waves (GW) by pulsars.
The  observed quantities are similar in both cases, and the power spectrum of pulsar frequency change caused by Doppler or Shapiro effects is red
 similar to gravitational waves. That is, there is an excess power at low frequencies,
 or long timescale correlations in residuals. 
The gravitational wave effect on pulsar timing is also in the
$nHz$ frequency range \cite{Jenet2005,Hobbs2009}, similar to the substructure effect we are considering here.

{In particular, the frequency change due to gravitational waves is
roughly $\sim h_{ij}$, the amplitude of gravitational waves, which
allows us to directly translate constraints on $h_{ij}$, to
constraints on $\delta \nu /\nu$. Moreover, similar to the
characteristic quadrupolar pattern that gravitational waves induce
in pulsar timing residuals ({\em e.g.,} \cite{Hobbs2009}), the
Doppler effect induces a {\em dipolar} pattern in the sky, which can
be used to distinguish it from intrinsic changes in individual
pulsars.}

More specifically, the frequency shift due to the combination of Doppler effect, gravitational waves, and intrinsic effects is given by:
\beq
\frac{\delta \nu(t)}{\nu}|_a = I_a(t) + {\bf \hat{n}}_a\cdot{\bf v}(t) + \frac{{\bf \hat{n}}_a\cdot{\bf h}({\bf k},t)\cdot{\bf \hat{n}}_a}{1+{\bf \hat{k}}\cdot{\bf \hat{n}}_a},
\eeq
where ${\bf \hat{n}}_a$ is the unit vector along the direction of pulsar $a$, $I_a(t)$ is the frequency shift intrinsic to the pulsar, ${\bf v}(t)$ is the earth's velocity, and ${\bf h}({\bf k},t)$ is the amplitude of a gravitational wave with wave-vector ${\bf k}$. The cross-power spectrum of frequency-change between different pulsars is given by
\bea
&&P_{\delta \nu \over \nu}(\omega)|_{ab} =  P_{\delta \nu \over \nu}(\omega)|_{int.} \delta(x_{ab}) \nonumber\\&+&P_{\delta \nu \over \nu}(\omega)|_{Doppler} (1-2x_{ab})+ P(\omega)|_{grav.} c(x_{ab}),
\eea
where
\beq
x_{ab}\equiv(1-{\bf \hat{n}}_a\cdot{\bf \hat{n}}_b)/2,
\eeq
and
\beq
c(x) \equiv \frac{3}{2}x\ln x-\frac{x}{4}+\frac{1}{2},
\eeq
is the expected correlation pattern of timing residuals for an isotropic stochastic gravitational wave background \cite{Hobbs2009}.  Therefore, pulsar timing cross-power spectra are affected by the intrinsic, Doppler and gravitational waves, $P_{\delta \nu \over \nu}(\omega)|_{int.}, P_{\delta \nu \over \nu}(\omega)|_{Doppler}$, and  $P(\omega)|_{grav.}$ with different angular dependences, which can be used to distinguish these effects.

\begin{figure}
\includegraphics[width=\linewidth]{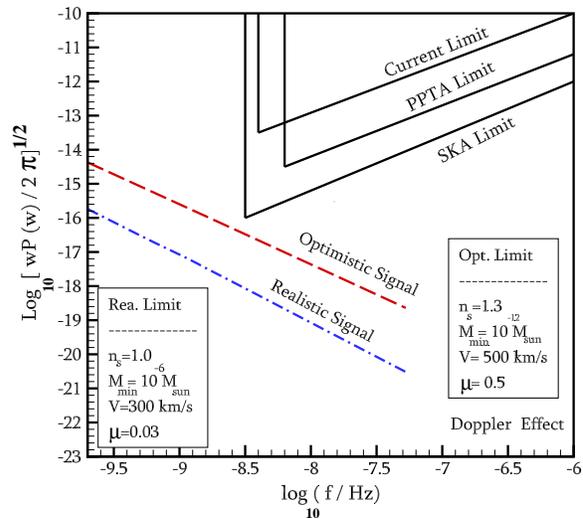}
\caption{Pulsar residual power spectrum as a function of the
frequency for Doppler effect for realistic and optimistic signals
(see text for definition of realistic and optimistic parameters).
The limits from current and future experiments are also shown.}
\label{H-D-op-6}
\end{figure}

 In Fig. (\ref{H-D-op-6}) we plot the realistic
and optimistic predictions for detection of $h_{p}$, which is similar to
the gravitational wave dimensionless strain,
and compare it with the current observational limits from a pulsar
timing array \cite{Jenet2006} considering the sensitivity limit  for time residuals of observed millisecond pulsars obtained via (see the appendix of \cite{Sesana2008} for details)
\begin{equation}
h^{lim}_{p}\propto\frac{\delta t_{rms}f}{N^{1/2}_{p}(T\Delta f)^{1/4}},
\end{equation}
where $h^{lim}_{p}$ is the sensitivity limit of detectors, $\delta
t_{rms}=\sqrt{\langle\delta t^2 \rangle}$ is the root mean square
value of the timing residuals, $\Delta f$ is the frequency bandwidth
of search, $N_{p}$ is the number of pulsars, and $T$ is the time
span of observation. The pulsar timing array sensitivity is scaled
with frequency as $h^{lim}_{p}\propto f$ and reaches a minimum at a
detectable frequency of $f\sim 1/T$. This produces the wedge-like
sensitivity limit curves in Fig.~(\ref{H-D-op-6}). The sensitivity
limit is also proportional to $\delta t_{rms}$, improving as the
precision of pulsar timing residuals  detection is increased. By
increasing the observational time, we increase the sensitivity and
also the span of frequency.

We also plot the predicted sensitivity of  Parkers pulsar timing
array (PPTA) \cite{Manchester2010} and the square kilometer array
(SKA) \cite{Smiths2008} for $h_{p}$. The  upper bounds for future
PPTA and SKA experiments are obtained from the detectable time
residual correlation of simulated pulsars with  consideration of all
instrumental, calibration and observational errors (such as pulsar
intrinsic period changes  and glitches)\cite{Hobbs2010}. For example
the PPTA bound is obtained by considering 20 radio pulsars for 5
year with $\delta t_{rms}=100 ns$ which provides a peak sensitivity
of $h^{lim}_{p}\approx 2\times 10^{-15}$ at $f\approx 7\times
10^{-9}$. \green{For SKA, with the same number of pulsars, the
sensitivity is improved by increasing the span of observation to 10
years with timing accuracy $\delta t_{rms}= 10 ns$, leading to a
constraint on the pulsar residual power spectrum of $\sim 1.6\times
10^{-16}$ at $f\approx 7\times 10^{-9}$ \cite{Sesana2008}.}

{ Finally, we study the effect of uncertainty in the models of
nonlinear structure formation on our results. In other words, how
much will our results depend on the choice of stable clustering
hypothesis? As we argued above,  stable clustering is the only known
physical prediction for the nonlinear power spectrum on very small
scales. Nevertheless, we can calculate $h_{p}$ for the nonlinear
power spectra of other clustering models in Fig.~(\ref{power-9}).
The relative magnitude of $h_{p}$ in two different models is
obtained from Eq.(\ref{powerdoppler}):
\begin{equation}\label{rel-hp}
\frac{h_{p}^{(m1)}}{h_{p}^{(m2)}}=\left[({\int
dk_{*}\frac{k_{*}^3}{k^4}P_{NL}^{(m1)}(k)})/({\int
dk_{*}\frac{k_{*}^3}{k^4}P_{NL}^{(m2)}(k)})\right]^{1/2},
\end{equation}
where superscript $(m1)$ and $(m2)$ indicate the models. In Fig.~
(\ref{fig:NL}), we plot $h_{p}$ for different models of
nonlinear structure formation by using the realistic parameters for the models.

\begin{figure}
\includegraphics[width=\linewidth]{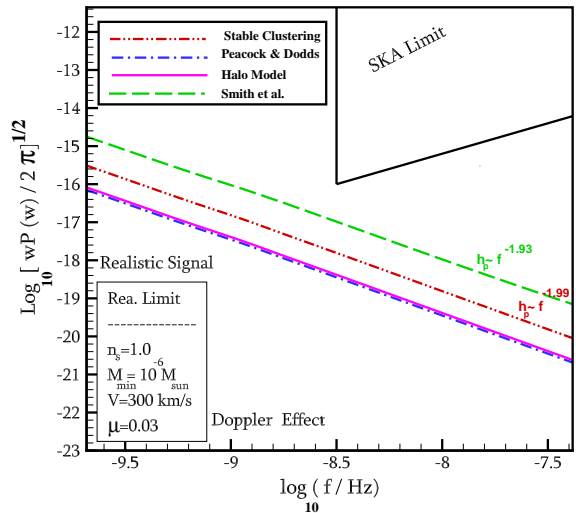}
\caption{Pulsar residual power spectrum as a function of the
frequency due to Doppler effect, for realistic signals in different
models of nonlinear structure formation.}
\label{fig:NL}
\end{figure}

An interesting point to notice is that different models of nonlinear
structure formation have (almost) the same frequency dependence,
$\omega P_{\delta\nu/\nu}\propto \omega ^{-4}$, as in stable
clustering. This is because of the moving screen approximation
$k_{x}v=\omega$ and $k_{z}\sim z^{-1}\ll k_{x},k_{y}$, which  is
applicable in the alternative models as well. On the other hand, the
main contribution of the integrals in Eq.~(\ref{rel-hp}) from the
nonlinear matter power spectrum comes when $k_{y}\sim \omega/v \sim
10 ^{-8}{\rm Hz}/300\mbox{ km}\sim 10^{9} \mbox{ Mpc}^{-1}$. In this
case $h_{p}^{m1}/h_{p}^{m2}$ reduces to the ratio of $P_{NL}$'s,
which is nearly independent of wavenumber (and thus frequency; see
Fig. \ref{power-9}) for relevant scales . {

Closer examination indicates that the frequency dependence of the
Smith et al. model is slightly shallower than the others
($h_{p}^{smith}\sim \omega ^{-1.93}$). This is due to the fact that
Smith et al. predict a much bluer spectrum on small scales (
Fig.\ref{power-9}), which is similar to the case of stable
clustering with higher power index (Fig.\ref{H-D-n-4}).

In summary, we find that the signature of the Doppler effect in
pulsar timing is largely independent of the nonlinear structure
formation model, which only introduces (a factor of a few)
uncertainty in the amplitude of timing residuals, $h_p$.}}
 Our results show that while current observations are unable to
detect the effect of dark matter substructure on pulsar timing,
error projections for the upcoming square kilometer array (SKA) are
only a factor of few higher than our optimistic predictions.

{In the end, we should note that, unlike the Doppler effect, the
Shapiro time delay does not have a coherent pattern on the sky, as
different lines of sight are largely uncorrelated. This makes it
much harder to distinguish Shapiro time delay from pulsar intrinsic
frequency changes.}



\section{Conclusions and Discussion}
\label{Section5} In this work, we studied the gravitational effect
of DM substructures on pulsar timing, through Doppler and Shapiro
(or ISW) effects.  We calculated the dimensionless power-spectrum of
a pulsar's frequency-change, which is related to the matter density
power-spectrum in the nonlinear regime. We used the stable
clustering hypothesis to extract the nonlinear matter
power-spectrum, and showed that the frequency-change is dominated by
the Doppler effect.  Next we varied the free parameters of the
model, which had the following effects on the dimensionless power,
$h_{p}$:
\begin{enumerate}
\item $h_p$ due to Doppler effect is linearly proportional to velocity of DM substructures.
\item The main contribution of DM substructures comes from the minimum
mass in DM hierarchy: as we increase the domain of integration over DM subhalo
masses, we  get more signal.
\item $h_{p}$ has a dependence on $\mu^{1/2}$, the fraction of particle pairs that remain bound in the stable clustering hypothesis.
\item $h_p$ due to the Shapiro effect scales as the square root of distances to pulsars, as it depends on the integrated gravitational effect over the line of sight.
\item  For larger primordial spectral index, $n_s$, the frequency  dependence of
$h_{p}$ is shallower, because the main contribution of $h_{p}$
comes from low masses, where the power is increased. However, for $n_s<1$, the frequency dependence becomes independent of $n_s$.

{\item The frequency dependence of $h_{p}$ is nearly independent of
nonlinear structure formation model, although its amplitude could
change by a factor of a few.}
\end{enumerate}
Finally, we compared the dimensionless power spectrum of pulsar
frequency change, for realistic and optimistic sets of parameters,
with current and future pulsar timing experiments, designed for
detection of gravitational waves. Our results show that our
optimistic estimate of the $h_{p}$ signal is only a  factor of a few
smaller than the sensitivity of the planned square kilometer array
(SKA), making this method a potentially promising avenue for the
detection of DM substructure on very small scales. While this may
sound too futuristic, it is worth noting that more dedicated pulsar
timing follow-ups of pulsars discovered by SKA, as well as better
noise removal techniques for ISM contamination of timing signals
({\em e.g.,} \cite{Demorest:2011hs}) will be able to potentially
push down the noise below our conservative forecasts.

We should further note that, as can be seen in Fig. (\ref{fig:NL}), if this signal is ever detected, there will be degeneracies
between parameters that quantify the nature of DM and those of structure formation (in both linear and nonlinear
regimes). Therefore, further study into the nature and properties of the signal (or independent observables)
will be necessary to disentangle these degeneracies.

{While our paper lays the groundwork for future statistical detection of dark matter substructure through pulsar timing, many practical challenges and theoretical uncertainties remain. Here we point out two, along with potential resolutions:}

{First, it is important to note that the observed Doppler effect in
pulsar timing depends on the total gravitational acceleration, which
can be contributed by nearby stars/planets, in addition to local
dark matter substructure. However,  the gravitational pull of
stars/planets on Earth can be calculated by knowing their masses and
positions around Earth, and thus, in principle, can be computed and
corrected for ({\em e.g.,} \cite{2010ApJ...720L.201C}). Similar
effects on the acceleration of pulsars will be uncorrelated for
different pulsars, and thus can be distinguished from Earth's
acceleration.}

{A second concern is the possible non-Gaussianity of the signal. For
example, microlensing events due to stars in the Galactic halo could
lead to large magnifications, but have very small optical depth, and
thus happen rarely. Therefore, the power spectrum gives a very
incomplete description of the observables in microlensing events.
However, in contrast to magnification events that trace projected
density, the gravitational effects on pulsar timing that we discuss
here trace the integrated potential, which is much more smooth.
Moreover, the small CDM substructure is much more diffuse than
stars, which further reduces the skewness of the signal. Therefore,
unlike microlensing events, the observed signal is likely to be
contributed by a variety of structures on different scales  ({\em
e.g.} Fig. \ref{H-D-M-3}) with no sharp boundaries. This is why we
expect a close to Gaussian signal, simply based on the central limit
theorem, which suggests that the power spectrum might provide
adequate statistical description of these effects.}

\begin{acknowledgments}
We would like to thank  Latham Boyle, Adrienne Erickcek, Sohrab Rahvar, and Ethan Siegal for discussion and valuable comments.
SB thanks the Perimeter Institute for their kind hospitality
during a visit where part of this work began.
NA is supported by Perimeter Institute (PI) for Theoretical Physics and Natural Sciences and Engineering Research Council of Canada (NSERC).  Research at PI is supported by the Government of Canada through Industry Canada and by the Province of Ontario through the Ministry of Research \& Innovation.
\end{acknowledgments}

\appendix
\section{Statistics of $\sigma_{z}$ for stability of Pulsars and
White noise calculation} \label{app1}
 The ability of a pulsar timing array to detect any delay in the received
pulses to measure the dark matter halos substructures depends on the
pulsar timing stability. Timing stability is related to how long
the rms of timing residuals can be kept small, from which we can
estimate the potential to detect Doppler and Shapiro effects. Statistical artifacts
such as a large gap in data sampling, or a large variation in error-bar
size, may prevent a reliable power spectrum of pulsar timing data.
An alternative approach is $\sigma_{z}$ statistics, as described by
e.g. Matsakis et al. \cite{Matsakis1997}:
\begin{equation} \label{sigmaz}
\sigma_{z}(\tau)=\frac{\tau^2}{2\sqrt{5}}\langle c_{3}^2\rangle^{1/2},
\end{equation}
where $\langle\rangle$ denotes the average over subsets of the pulsar timing data, and $c_{3}$ is determined from a polynomial fit
\begin{equation}
c_{0}+c_{1}(t-t_{0})+c_{2}(t-t_{0})^2+c_{3}(t-t_{0})^3
\end{equation}
to timing residuals for each subset, and $\tau$ is the length of the subsets.
In order to connect our theoretical calculations to the observed pulsar time residuals we should find a relation between
$\sigma_{z}$ and the calculated power spectrum.
From the polynomial fit to the  timing residuals we find that
\begin{equation}
c_{3}\simeq\frac{1}{6}\frac{d}{d\tau}\Delta\ddot{t}\mid_{s}\simeq\frac{1}{6}\frac{d}{d\tau}{\dot{\frac{\delta\nu}{\nu}}},
\end{equation}
where we assume that the fitting procedure depends on  $\Delta t
\mid_{s}$, which is coarse grained on the scale of $\tau$. The
correlation function $c_{3}$ can be written as
\begin{equation}\label{c3}
\langle c^{2}_{3}\rangle=\frac{1}{18}\{\langle{(\dot{\frac{\delta\nu}{\nu}})}^2\rangle-\langle\dot{(\frac{\delta\nu}{\nu})}|_{{\i}}\dot{(\frac{\delta\nu}{\nu})}|_{{\i}{\i}}\rangle\}.
\end{equation}
Now, using Eqs.(\ref{EqPS},\ref{sigmaz},\ref{c3})
we obtain
\begin{equation}
\sigma_{z}(\tau)\simeq\frac{\tau}{6\sqrt{5}}\left\{\int_{0}^{\frac{1}{\tau}}\frac{d\omega}{2\pi}\omega^2P(\omega)[1-\cos(\omega\tau)]\right\}^{1/2}.
\end{equation}
In order to find the white noise corresponding to pulsar timing
we derive the relation of the dimensionless power spectrum of pulsars
with the sampling time and the uncertainty in the pulsar timing
measurement. The cross correlation of time residuals of pulsar
timing is related to the accuracy of measurement $t_{a}$ as
\begin{equation}
\langle \delta t(t_{1})\delta
t(t_{2})\rangle=(t_{a})^2\delta_{t_{1}t_{2}},
\end{equation}
where $\delta_{t_{1}t_{2}}$ is the Kronecker delta and $\delta {t}$
is the time residual of pulsar timing related to frequency change as
\begin{equation}
\delta t=\int \frac{\delta\nu}{\nu} dt
\end{equation}
The correlation of timing residuals can be approximated in the time
span of $\tau$, which is the period of sampling as:
\begin{equation}
\langle \delta t(t_{1})\delta
t(t_{2})\rangle\simeq(t_{a})^2\tau\delta(t_{1}-t_{2})
\end{equation}
Now the power spectrum of time residuals is obtained as
\begin{equation}
P_{\delta t}(\omega)=\int e^{-i\omega t}\langle \delta
t(t_{1})\delta t(t_{2})\rangle dt =\tau t_{a}^2,
\end{equation}
which yields the dimensionless power spectrum,
\begin{equation}\label{WN}
h_{p}=\left[\frac{1}{2\pi}\omega
P_{\frac{\delta\nu}{\nu}}(\omega)\right]^{1/2}=\frac{\sqrt{\tau}}{\sqrt{2\pi}}\omega^{3/2}t_{a}.
\end{equation}
To find the white noise lines in Fig.(\ref{H-Sh-D}), we set the sampling time of pulsar timing $\tau$ to be 2 weeks and
the accuracy of pulsar timing, $t_a$, to be $100 ns$ and $1 \mu s$.


\begin{thebibliography} {50}

\bibitem{WMAP}
N. Jarosik {\it{et al.}}, arXiv:1001.4744 (2010).

\bibitem{Roos2010}
M. Roos, Dark Matter: The evidence from astronomy, astrophysics and cosmology.
arXiv:1001.0316 (2010).

\bibitem{kris2006}
A. Loeb and M. Zaldarriaga, Phys. Rev. D {\bf{71}}, 103520 (2005);
S. Profumo, K. Sigurdson and M. Kamionkowski, Phys. Rev. Lett. {\bf{97}}, 031301 (2006);
G.D. Martinez, J.S. Bullock, M. Kaplinghat, L.E. Strigari and R. Trotta, JCAP {\bf{0906}},
014 (2009).

\bibitem{Frail1994}.
D.A. Frail {\it {et al.}}, Astrophysical J., Vol. 1, {\bf{436}}, no.1, 144 (1994).

\bibitem{Shapiro1964}
I.I. Shapiro, Phys. Rev. Letters {\bf{13}}, 789 (1964).

\bibitem{Siegel2007}
E.R. Siegel, M.P. Hertzberg and J.N. Fry, MNRAS,  382, Issue 2, 879 (2007).

\bibitem{Seto2007}
N. Seto and A. Cooray, Astrophys. J. {\bf{659}}, 33 (2007).

\bibitem{Pshirkov2008}
M. Pshirkov, A. Tuntsov and K.A. Postnov, Phys. Rev. Lett. {\bf{101}}, 261101 (2008).

\bibitem{Ishiyama2010}
T. Ishiyama, J. Makino and T. Ebisuzaki, Astrophys. J. Lett. V. 273, I. 2, 195 (2010).

\bibitem{Longo1988}
M.J. Longo, Phys. Rev. Lett. {\bf{60}}, 173 (1988).

\bibitem{Larchenkova2006}
T.I. Larchenkova and S.M. Kopeikin, Astronomy Letters, Vol. 32, {\bf{1}}, 18 (2006).

\bibitem{Ohnishi1995}
K. Ohnishi, M. Hosokawa, T. Fukushima and M. Takeuti, Astrophys. J. {\bf{448}},
271 (1995); T.I. Larchenkova and O.V. Doroshenko, Astron.  Astrophys., {\bf{297}}, 607 (1995);
M.A. Walker, Publ. Astron. Soc. of Australia, V. 13, no. 3, 236 (1996);
M. Hosokawa, K. Ohnishi, and T. Fukushima, Astron. Astrophys. {\bf{351}}, 393 (1999);
M.S. Pshirkov, M.V.  Sazhin and Yu.P. Ilyasov,  Astronomy Letters, V. 34, I. 6, 397 (2008).


\bibitem{Erick2010}
A. Erickcek and N.M. Law, arXiv: 1007.4228 (2010).

\bibitem{Peebles77}
M. Davis and P.J.E. Peebles, ApJS {\bf{34}}, 425 (1977).

\bibitem{Hamilton1977}
A.J.S. Hamilton, A. Matthews, P. Kumar and E. Lu, Astrophys. J.
{\bf{374}}, L1 (1991); B. Jain, H.J. Mo  and S.D.M. White, Mon. Not.
Roy. Astron. Soc. {\bf{276}}, L25 (1995).



\bibitem{Peacock1996}
J.A. Peacock and S.J. Dodds, Mon. Not. Roy. Astron. Soc.
{\bf{280}}, L19 (1996).


\bibitem{Afshordi2009}
N. Afshordi, R. Mohayaee and E. Bertschinger, Phys. Rev. D {\bf{81}}, 101301 (2010).



\bibitem{Verb2009}
J.P.W. Verbiest {\it{et al.}}, MNARS, Volume 400, Issue 2, 951 (2009).


\bibitem{Jenet2006}
F.A. Jenet {\it {et al.}}, ApJ. {\bf{653}}, 1571 (2006).


\bibitem{Jenet2005}
F.A. Jenet, G.B. Hobbs, K.J. Lee and R.N. Manchester, Astrophys. J. {\bf{625}}, 123 (2005).

\bibitem{Sachs1967}
R.K. Sachs, and A.M. Wolfe, Astrophys. J. 147, 73 (1967).

\bibitem{Gunn72}
J.E. Gunn and J.R. Gott, Astrophys. J. {\bf {176}}, 1 (1972).

\bibitem{Bardeen1986}
J.M. Bardeen, J.R. Bond, N. Kaiser and A.S. Szalay ,   Astrophys. J., {\bf{304}},
15 (1986).

\bibitem{Cooray:2002dia}
A. Cooray and R.K. Sheth, Phys. Rept. {\bf 372}, 1 (2002).

\bibitem{Smith2003}
R.E. Smith {\it {et al.}}, MNRAS , Volume 341, Issue 4, 1311 (2003).

\bibitem{Hobbs2009}
G. Hobbs, F.A. Jenet, K.J. Lee, J.P.W. Verbiest, D. Yardley, R. Manchester, A. Lommen, W. Coles, R. Edwards and
C. Shettigara,  MNRAS, {\bf{394}}, 1945 (2009).

\bibitem{Sesana2008}
A. Sesana, A. Vecchio and C.N. Colacino, MNARS, {\bf{390}}, 192 (2008)


\bibitem{Manchester2010}
R. N. Manchester, arXiv: 1004.3602 (2010);
G.B. Hobbs {\it{et al.}}, PASA {\bf{26}}, 103 (2009);
J.P.W. Verbiest, {\it{et al.}}, Class. and Quant. Grav. {\bf{27}}(8), 084, 015 (2010).

\bibitem{Smiths2008}
R. Smiths, {\it{et al.}}, Pulsar searches and timing with SKA, arXiv:0811.0211 (2008).

\bibitem{Hobbs2010}
G. Hobbs, Pulsars as Gravitational Wave detectors, arXiv: 1006.3969 (2010).

\bibitem{Demorest:2011hs}
  P.~Demorest,
  arXiv:1106.3345 [astro-ph.IM].

\bibitem{2010ApJ...720L.201C} Champion, D.~J., et al.\ 2010,  Astrophys. J., 720, L201

\bibitem{Matsakis1997}
D.N. Matsakis, J.H. Taylor and T.M. Eubanks, AA, {\bf{326}}, 924 (1997).

\end{thebibliography}
\end{document}